\documentclass[nocompress]{spie}  %>>> to avoid compression of citations
\usepackage[]{xspace,amsmath,graphicx}
\usepackage[caption=false]{subfig}
\newcommand{\FIG}[3]{\includegraphics[width=#1\linewidth,draft=#2]{#3.eps}}
\newcommand{\FIGH}[3]{\includegraphics[height=#1cm,draft=#2]{#3.eps}}

\newcommand{\mus}{\mbox{{\usefont{U}{eur}{m}{n}{\char22}}s}\xspace}
\newcommand{\loD}{\mbox{$\lambda/D$}\xspace}
\newcommand{\e}[1]{10^{#1}}
\newcommand{\E}[1]{\times10^{#1}}
\newcommand{\Coro}{Coronagraph\xspace}

\newcommand{\coro}{coronagraph\xspace}
\newcommand{\coros}{coronagraphs\xspace}

\graphicspath{{D:/Documents/2_EXCEDE/DATA/}{D:/Documents/2_EXCEDE/REPORTS/}}

\title{Experimental study of a low-order wavefront sensor for the high-contrast coronagraphic imager EXCEDE} 

\author{Julien~Lozi\supit{a}, Ruslan~Belikov\supit{b}, Glenn~Schneider\supit{a}, Olivier~Guyon\supit{a}, Eugene~Pluzhnik\supit{b,c}, Sandrine~J.~Thomas\supit{b,c}, Frantz~Martinache\supit{d}\skiplinehalf
\supit{a}University of Arizona, 1401 E University Blvd, Tucson, AZ 85721, USA; \\
\supit{b}NASA Ames Research Center, Moffett Field, CA 94035, USA; \\
\supit{c}UARC/NASA Ames, P.O. Box 7, Moffett Field, CA 94035, USA; \\
\supit{d}Subaru Telescope, National Astronomical Observatory of Japan, 650 North A'ohoku Place, Hilo, HI 96720, USA
}

\authorinfo{Further author information: (Send correspondence to J.L.)\\ J.L.: E-mail: jlozi@email.arizona.edu, Telephone: 1 650 604 6549}
\begin{document} 
  \maketitle 

%%%%%%%%%%%%%%%%%%%%%%%%%%%%%%%%%%%%%%%%%%%%%%%%%%%%%%%%%%%%% 
\begin{abstract}
\end{abstract}

The mission EXCEDE (EXoplanetary Circumstellar Environments and Disk Explorer), selected by NASA for technology development, is designed to study the formation, evolution and architectures of exoplanetary systems and characterize circumstellar environments into stellar habitable zones. It is composed of a 0.7~m telescope equipped with a Phase-Induced Amplitude Apodization \Coro (PIAA-C) and a 2000-element MEMS deformable mirror, capable of raw contrasts of $\e{-6}$ at 1.2~\loD and $\e{-7}$ above 2~\loD. One of the key challenges to achieve those contrasts is to remove low-order aberrations, using a Low-Order WaveFront Sensor (LOWFS). An experiment simulating the starlight suppression system is currently developed at NASA Ames Research Center, and includes a LOWFS controlling tip/tilt modes in real time at 500~Hz. The LOWFS allowed us to reduce the tip/tilt disturbances to $\e{-3}$~\loD rms, enhancing the previous contrast by a decade, to $8\E{-7}$ between 1.2 and 2~\loD. A Linear Quadratic Gaussian (LQG) controller is currently implemented to improve even more that result by reducing residual vibrations. This testbed shows that a good knowledge of the low-order disturbances is a key asset for high contrast imaging, whether for real-time control or for post processing.

%>>>> Include a list of keywords after the abstract 


%%%%%%%%%%%%%%%%%%%%%%%%%%%%%%%%%%%%%%%%%%%%%%%%%%%%%%%%%%%%%

\section{INTRODUCTION}
\label{sec:intro}

In the next generation of ground and space \coros, stability of the instrument is a key issue to obtain good contrasts at small Inner Working Angles (IWA). Space telescopes are affected by vibrations, due to reaction wheels for example, as well as pointing stability, while ground telescopes are mostly affected by turbulence and vibrations. Therefore, if we want to image planets and stellar environments at 1~AU, \coros will have to be equipped with Low-Order Wavefront Sensors (LOWFS), capable of measuring and correcting low-order aberrations.

In the context of the EXCEDE (EXoplanetary Circumstellar Environments and Disk Explorer) mission\cite{Guyon12}, we are currently testing its starlight suppression system (the \coro with the wavefront correction and the LOWFS) in air, at the Ames Coronagraphic Experiment (ACE) at NASA Ames Research Center (Moffett Field, CA) and soon in vacuum at Lockheed Martin (Palo Alto, CA). A description of the goals of EXCEDE and ACE is presented during this conference\cite{Belikov13}, as well as the setup and the results of the wavefront control\cite{Thomas13}.

In this paper, we will present the experimental analysis and the results we obtained on the LOWFS we implemented in the ACE laboratory. Section~\ref{sec:TheLowOrderWavefrontSensor} provides a description of the principle of the LOWFS, as well as the hardware and software architecture. Then Sec.~\ref{sec:CalibrationProcedures} describes the different calibration procedures of the system. Finally, section.~\ref{sec:ExperimentalResults} presents the performances of the LOWFS and analyzes its sensitivity to different parameters.

%%%%%%%%%%%%%%%%%%%%%%%%%%%%%%%%%%%%%%%%%%%%%%%%%%%%%%%%%%%%%

\section{The Low-Order Wavefront Sensor}
\label{sec:TheLowOrderWavefrontSensor}

%%-----------------------------------------------------------

\subsection{Principle}
\label{sec:Principle}

The coronagraphic low-order wavefront sensor, developed by O.Guyon, is well described in\cite{Guyon09}. It uses a defocused image of the light rejected by the focal plane occulter. Images are then compared to a reference, and decomposed on a base of orthogonal modes. It was developed and successfully tested for PIAA \coros\cite{Guyon09,Vogt10}, especially the SCExAO instrument of the Subaru Telescope\cite{Martinache09}. This type of LOWFS is not exclusively designed for PIAA \coros, it can also be implemented on any type of \coro that uses a focal plane occulter.

A similar design was also developed for \coros with complex masks using the light rejected by the Lyot stop, such as PIAA Complex Mask \Coro (PIAACMC)\cite{Guyon10}, the vortex or the four quadrant phase mask (FQPM). It has been already tested on sky on SCExAO\cite{Singh13}. The real-time measurements can also be used during post-processing, to remove a maximum of residual starlight. This technique was successfully tested in a lab experiment\cite{Vogt11}.

For a PIAA \coro, the main mode to correct we have to correct are the pre-PIAA tip and tilt. Because of the shape of the PIAA optics, pre-PIAA tip/tilt is different from post-PIAA tip/tilt. In a first order, the first one can compensate the second one, but if the disturbances are too large, both has to be controlled.

%%-----------------------------------------------------------

\subsection{Hardware setup}
\label{sec:HardwareSetup}

\begin{figure}[b] \centering
  \subfloat[Photo of a 3-zone focal plane occulter.]
	{\label{fig:FPO}\FIGH{5}{false}{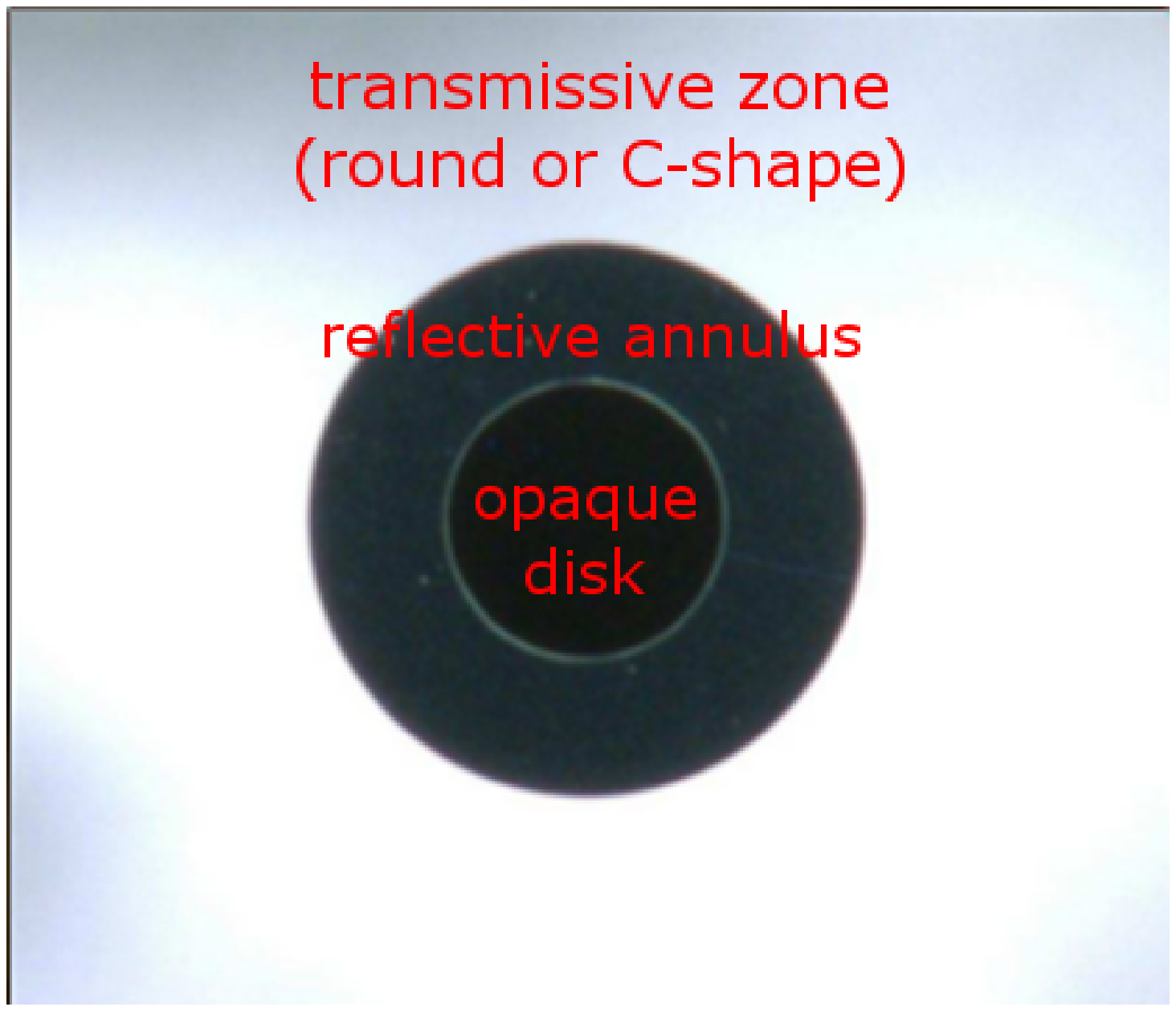}}\hspace{20pt}
  \subfloat[Camera used for the LOWFS.]
	{\label{fig:cam}\FIGH{5}{false}{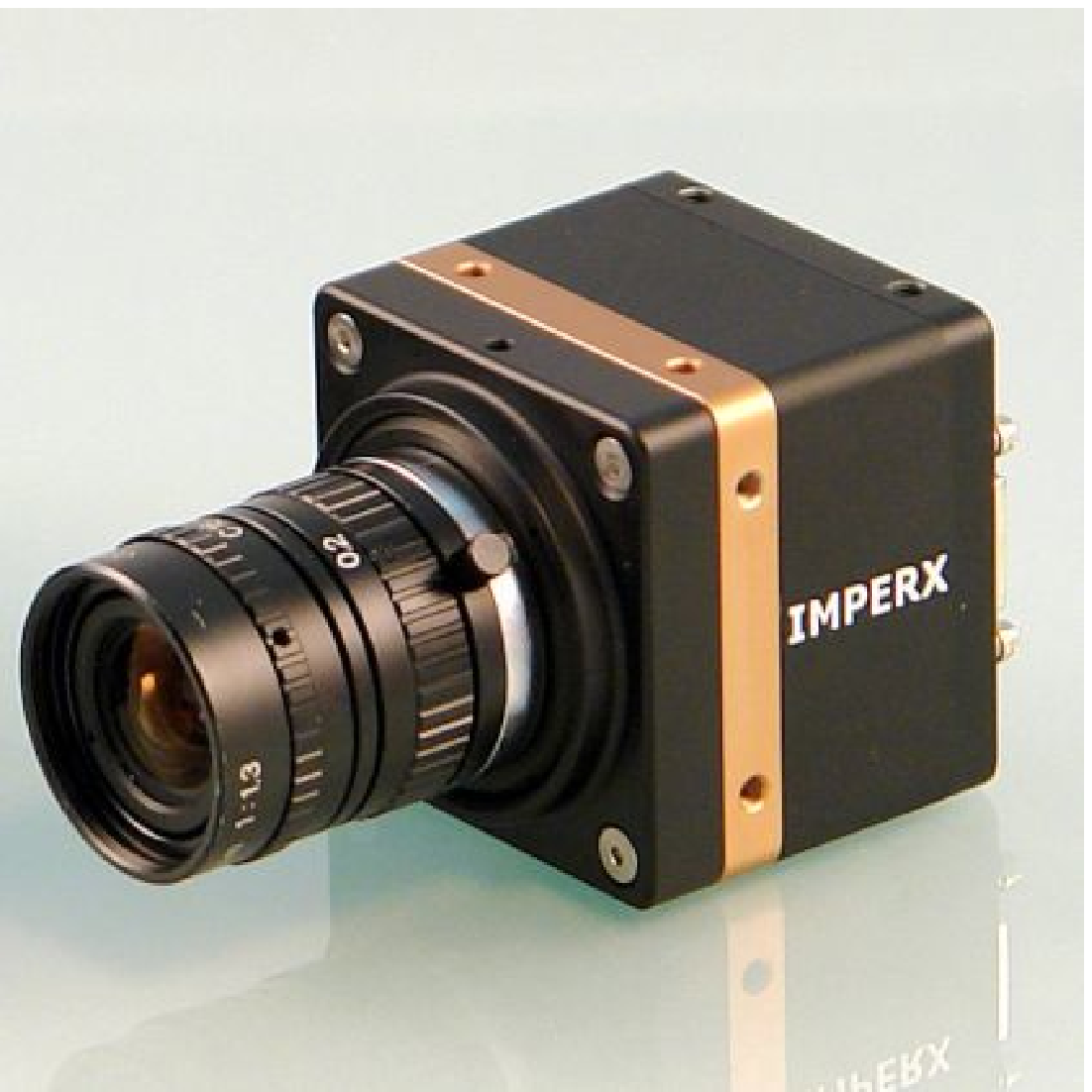}}
  \caption{Hardware composing the LOWFS.}
  \label{fig:hardware}
\end{figure}

The design of the LOWFS is simply composed of three elements:
\begin{itemize}
	\item A focal plane mask. At first, we used a fully reflective mask with a transmissive C-shape; now we are using a 3-zone mask as described in\cite{Guyon09}: a transmissive zone for the planet light, an opaque disk to block the core of the star, and a reflective annulus to keep the edges of the star (see Fig~\ref{fig:FPO}).
	\item A lens that collects the light from the star reflected by the focal plane mask, and sends it to a camera. This lens does not need to be of very good quality or achromatic.
	\item A fast camera that takes images of the starlight with a frequency high enough to correct eventual disturbances.
\end{itemize}

The camera we are using is an Imperx Bobcat ICL-B0610 with a 648x488 detector, up to 14~bits per pixel, with a frequency of 105~Hz in full frame (see Fig.~\ref{fig:cam}). A center mode can be enabled, which reads only 228 columns at the center of the detector. In that mode, the frame rate is 300~Hz, and it goes up to 1.8~kHz if we reduce the number of lines. In our application, a window of $50\times50$~pixels is large enough, the frame rate is then 1.1~kHz.

The images are grabbed by a real-time computer, a National Instrument PXI. It is used to perform fast calculations without jitters. Commands are then calculated and sent to actuators. In our case, we control the $x$- and $y$-axis translations of the source, corresponding to pre-PIAA tip/tilt, but the system is very flexible, and we can add other modes: focus, post-PIAA tip/tilt, astigmatism, etc.

%%-----------------------------------------------------------

\subsection{Software architecture}
\label{sec:SoftwareArchitecture}

The software architecture is made up of three different loops:
\begin{itemize}
	\item The Time-Critical Loop (TCL): this loop grabs the images from the camera, computes the command and sends it to the actuators.
	\item The Graphical User Interface (GUI): this loop gives a visualization of the controller state, and allows the user to change some parameters.
	\item The Non-Critical Loop (NCL): this loop transfers the parameters from the GUI to the TCL, and also transfers the data from the TCL to the GUI. 
\end{itemize}

\begin{figure} \centering
\FIG{0.5}{false}{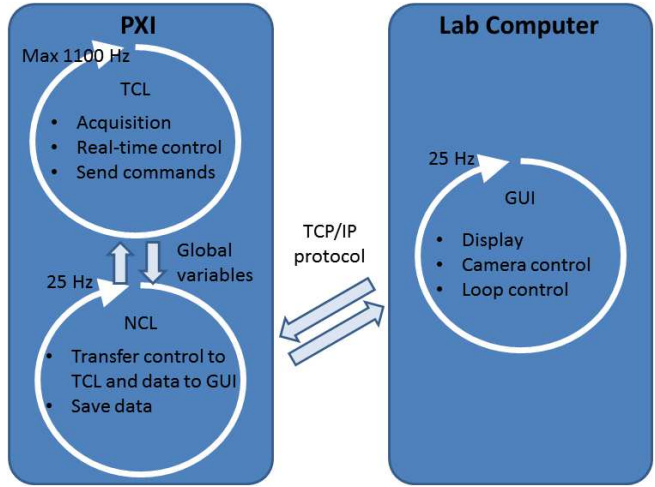}
\caption{Software architecture of the LOWFS.}
\label{fig:archi}
\end{figure}

Figure~\ref{fig:archi} summarizes the software architecture. The TCL and the NCL are running independently on the PXI, while the GUI is running on a standard lab computer. The two systems communicate through a TCP/IP protocol. All the time critical operations are made on the real-time computer, to avoid jitters induced by operations with a lower priority.

%%%%%%%%%%%%%%%%%%%%%%%%%%%%%%%%%%%%%%%%%%%%%%%%%%%%%%%%%%%%%

\section{Calibration procedures}
\label{sec:CalibrationProcedures}

%%-----------------------------------------------------------

\subsection{Calibration of the LOWFS}
\label{sec:CalibrationOfTheLOWFS}

The calibration of the LOWFS is simple, and it is made up of two steps.

First, we save a reference position for the system. Since we have a constraint about the inner working angle (IWA) ---~defined as the position where the throughput is exactly 50\%~--- and since the mask might not exactly match this IWA, usually the PSF is not exactly centered on the mask. So the reference image is usually not symmetrical.

\begin{figure} \centering
  \subfloat[Reference image after a calibration.]
	{\label{fig:refim}\FIG{0.3}{false}{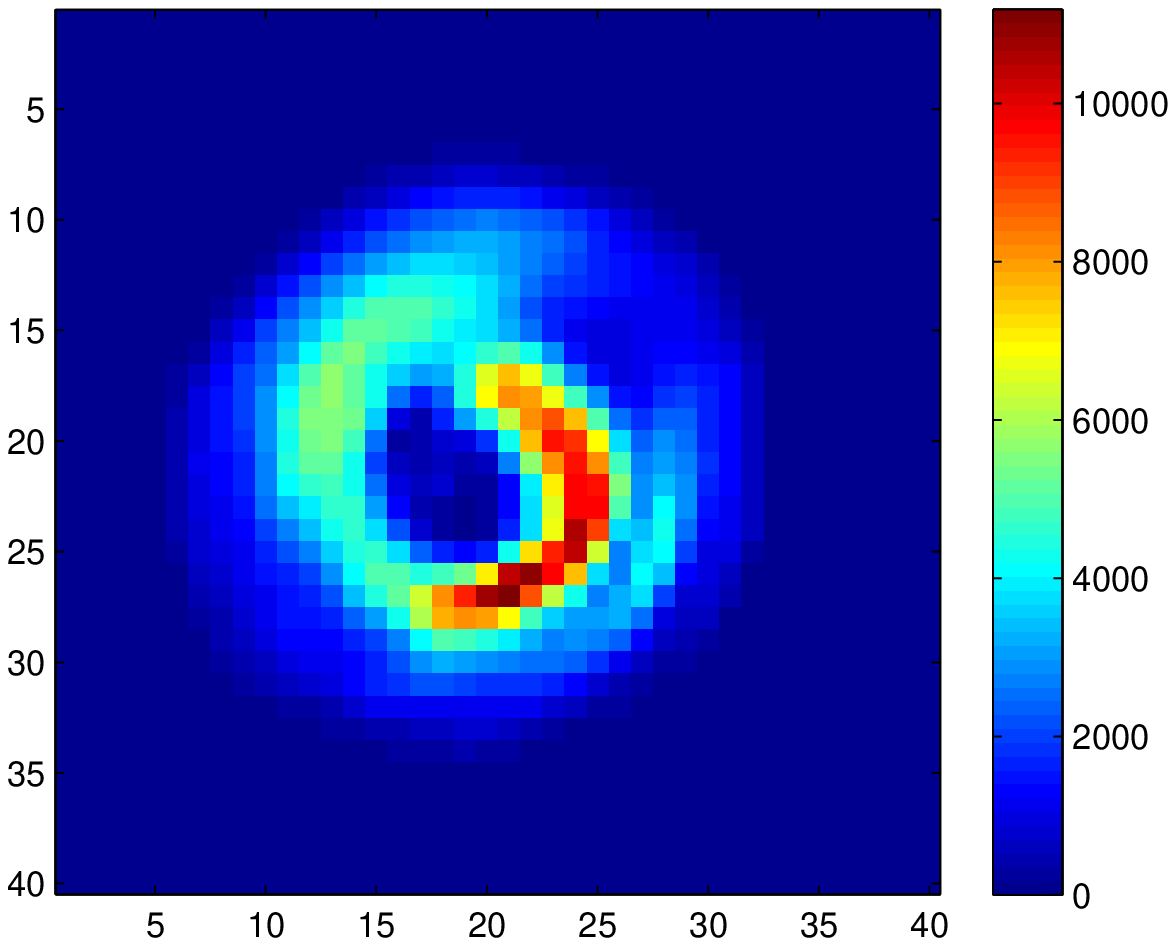}}\hspace{20pt}
  \subfloat[Signal corresponding to the difference between an image and the reference.]
	{\label{fig:signal}\FIG{0.3}{false}{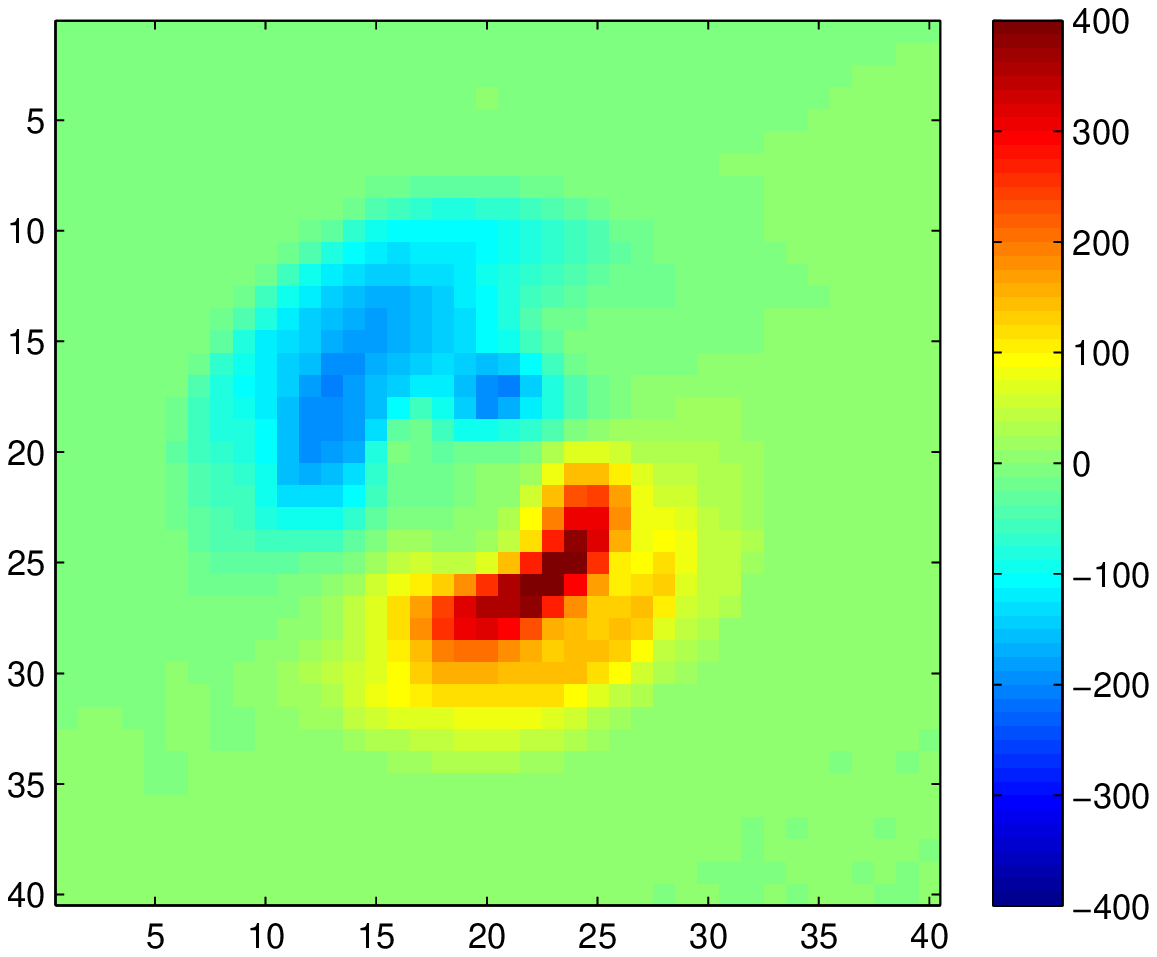}}
  \caption{Reference image and signal measured by the LOWFS.}
  \label{fig:reference}
\end{figure}

A typical reference image is presented in Fig.~\ref{fig:refim}. To obtain this image, we took an average of a few hundred images after putting the coronagraphic mask at the desired position. We can see that the image is slightly defocused, and the opaque disk is visible on the center. After obtaining this reference image, it is subtracted to every images coming from the camera. An example of the resulting difference is shown in Fig.~\ref{fig:signal}.

The LOWFS we developed provides two types of measurements. The first one is a classical measurement of the centroid of the image. The centroid is then compared to the centroid of the reference image. The second measurement uses the technique described in Sec.~\ref{sec:Principle}. For this measurement, there is a second step to the calibration procedure.

\begin{figure} \centering
  \subfloat[Mode 1 (pre-PIAA tip) of the LOWFS.]
	{\label{fig:mode1}\FIG{0.3}{false}{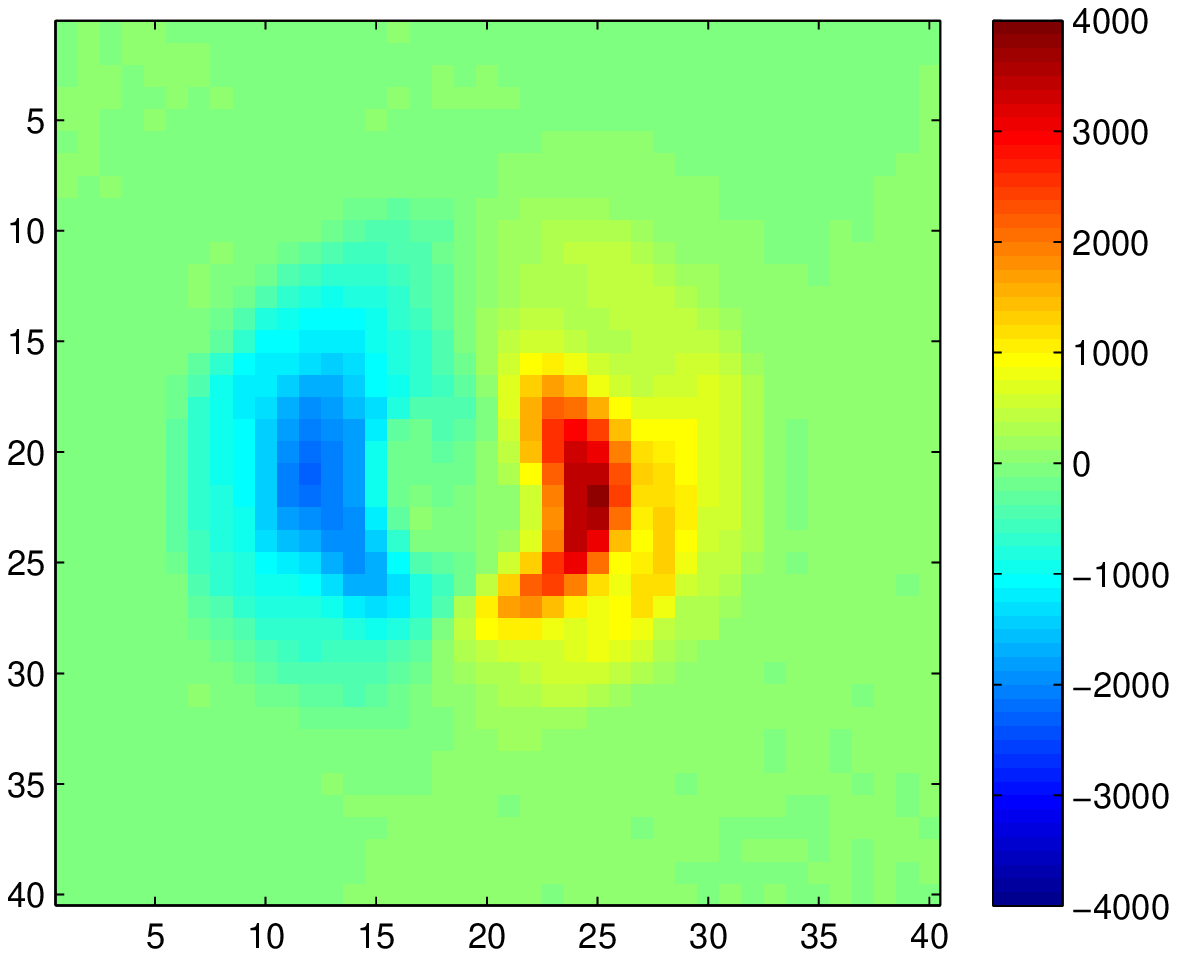}}\hspace{20pt}
  \subfloat[Mode 2 (pre-PIAA tilt) of the LOWFS.]
	{\label{fig:mode2}\FIG{0.3}{false}{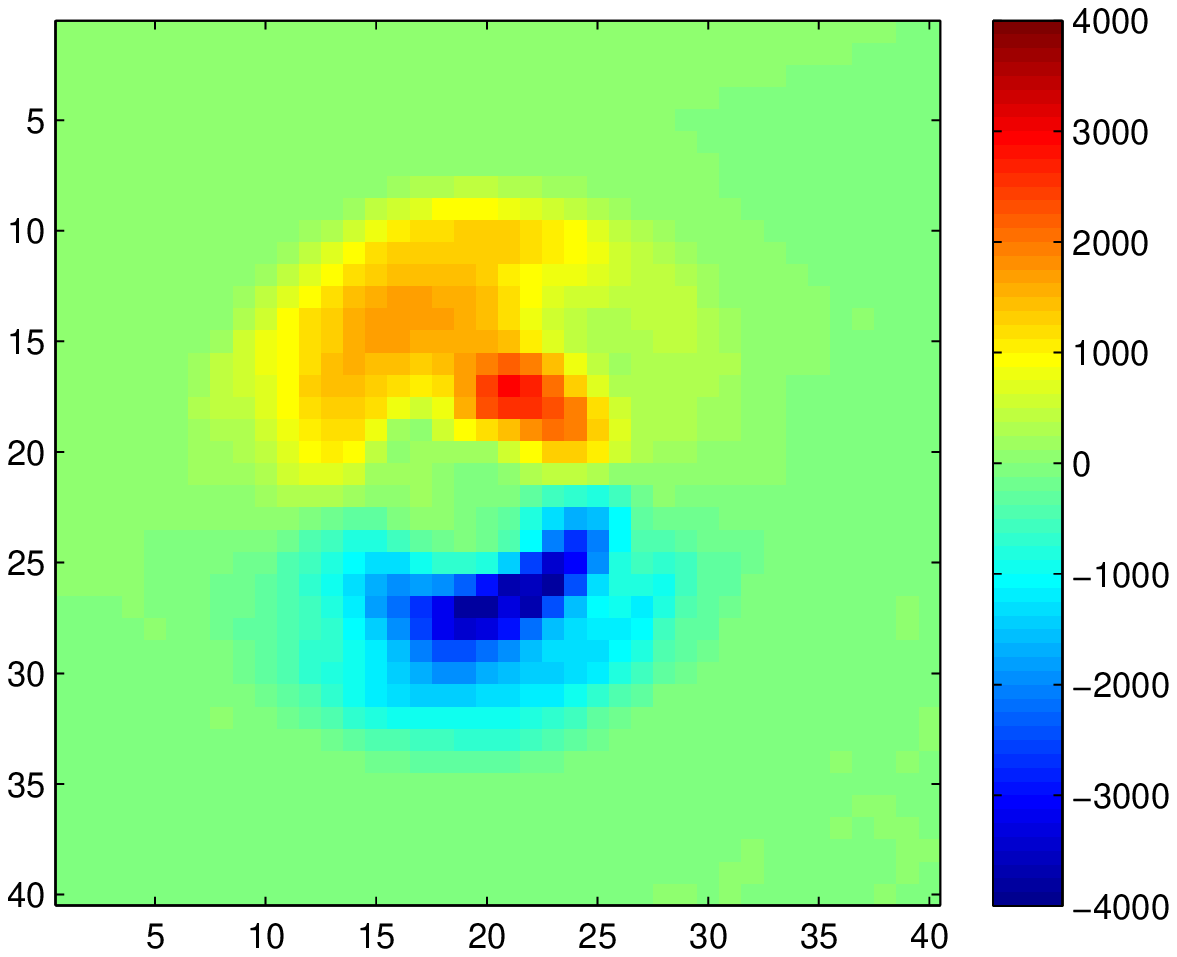}}
  \caption{Base of modes used in the LOWFS.}
  \label{fig:modes}
\end{figure}

In this step, we send known voltages to the different actuators we want to use ---~in this case some piezoelectric actuators on the $x$- and $y$-axis of the source~--- being careful to stay within the range of linearity. Then we measure the impact of each displacement on the signal. Figure~\ref{fig:modes} presents the pre-PIAA tip/tilt modes obtained with this calibration. Normally we should orthogonalize the modes to avoid cross-talk between the modes. In our case, we only correct pre-PIAA tip and tilt, so the modes are naturally almost orthogonal. It would be an essential step if we add the correction of post-PIAA tip/tilt for example, because the cross-talk between those modes and the pre-PIAA tip/tilt is important. This calibration is similar to the construction of the influence matrix in adaptive optics.

The measurement provided by the LOWFS is then the deconstruction of the signal presented in Fig.~\ref{fig:signal} into the modes presented in Fig~\ref{fig:modes}.

%%-----------------------------------------------------------

\subsection{Linearity and conversion into angle units on the sky}
\label{sec:ConversionIntoSkyAngleUnits}

The measurements provided by the program are not uniform: the centroid measurement is in camera pixels, while the LOWFS provides a tip/tilt measurement in units of modes, proportional to the voltage applied to the piezo actuators. So in order to compare the two methods, and for a better readability, we transform these units into \loD on the sky.

To perform this calibration, we scanned between $-1$ and 1~\loD on both axes, and we computed the best linear fit between $-0.1$ and 0.1~\loD. This method gives also a direct measurement of the linearity of the system. We performed this calibration with the two types of mask: the fully reflective mask and the 3-zone mask, as described in Sec.~\ref{sec:HardwareSetup}.

\begin{figure} \centering
  \subfloat[Linearity with the fully reflective mask.]
	{\label{fig:calibwodisk}\FIG{0.5}{false}{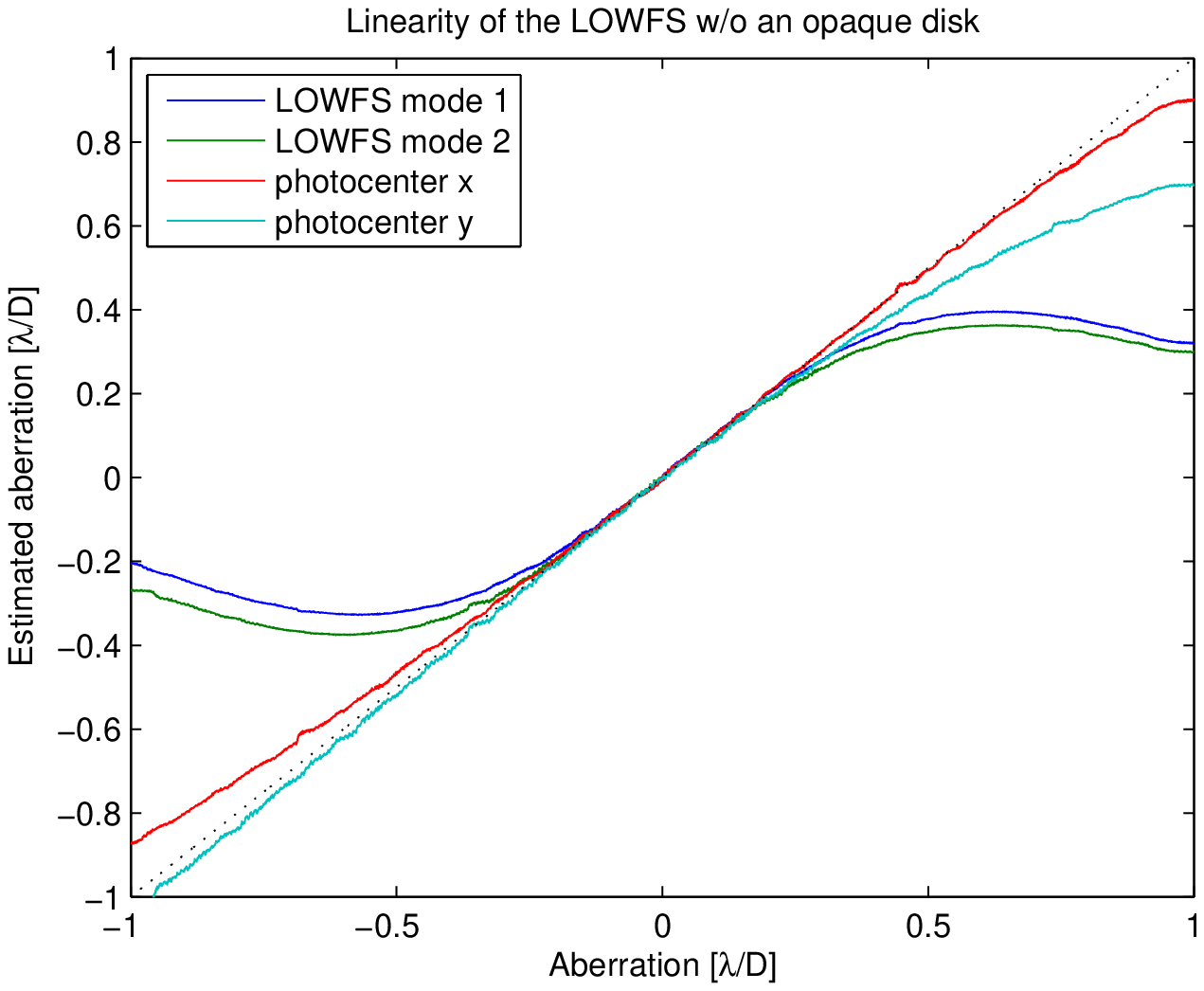}}
  \subfloat[Linearity with the 3-zone mask.]
	{\label{fig:calibwdisk}\FIG{0.5}{false}{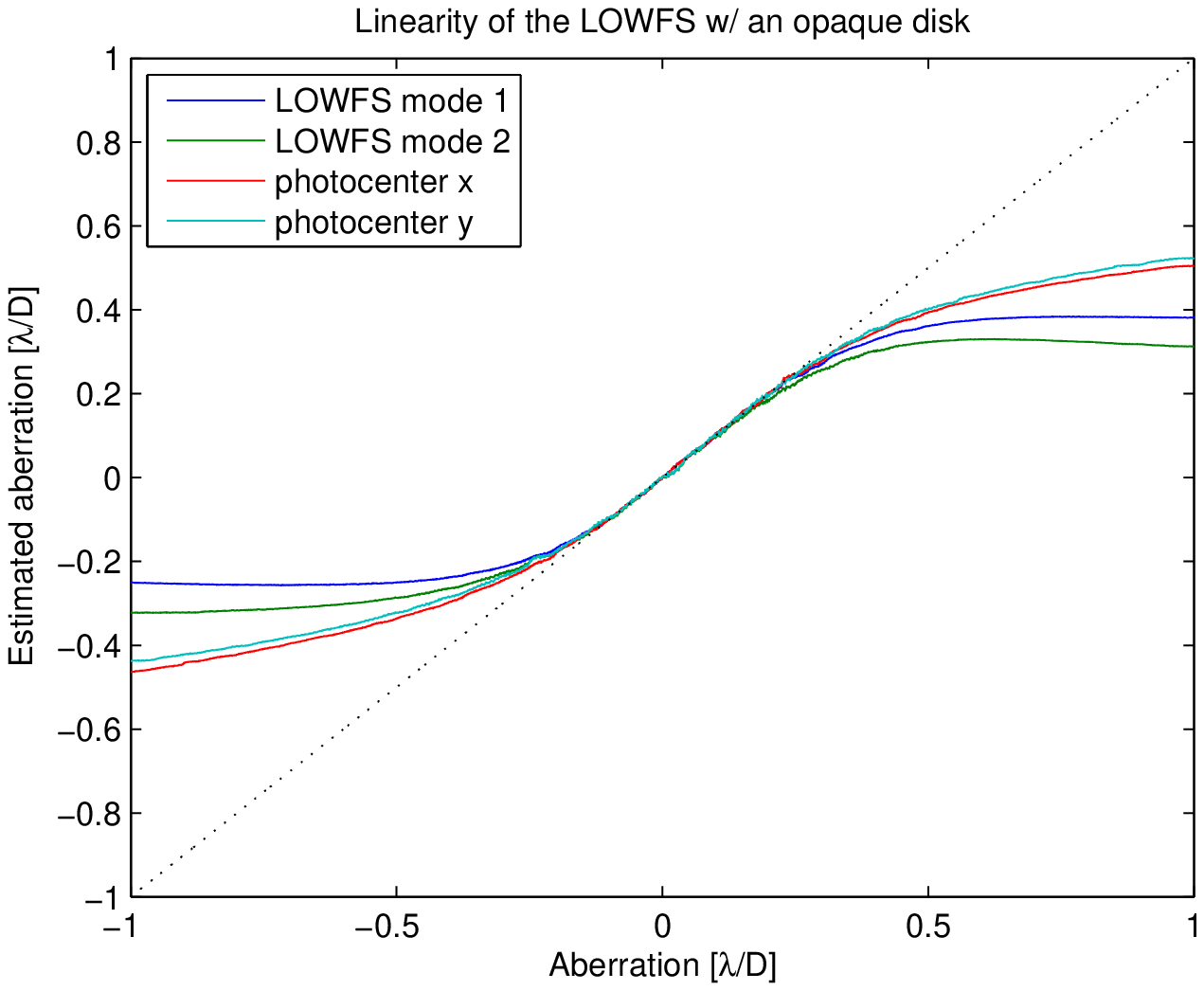}}
  \caption{Linearity of the LOWFS.}
  \label{fig:calib}
\end{figure}

The results are presented on Fig.~\ref{fig:calib}: Fig.~\ref{fig:calibwodisk} is the linearity of the system with the fully reflective mask, and Fig.~\ref{fig:calibwdisk} is the linearity with the 3-zone mask.

On Fig.~\ref{fig:calibwodisk}, we can see that the measurements are linear only on a small range of about 0.4~\loD for the LOWFS, and more than 1~\loD for the centroid. But while the centroid is monotonic over the whole range, the LOWFS measurement is monotonic only on a range of approximately 1.2~\loD: it corresponds to the position where the PSF does not overlap anymore with the reference PSF. We can also notice that the error in the centroid measurement of figure~\ref{fig:calibwodisk} is higher at 1~\loD than at $-1$~\loD: This is due to the asymmetric transmission of the C-shape focal plane mask.

On Fig.~\ref{fig:calibwdisk}, the linearity range is approx. the same for both the LOWFS and the centroid measurement, about 0.4~\loD. For this mask, the centroid is also monotonic over the whole range, while the LOWFS measurement has a plateau after $\pm0.5$~\loD.  We observe also a slight difference between modes 1 and 2 of the LOWFS, it is probably because the star is not perfectly centered on the mask.

So if we correct only pre-PIAA tip/tilt modes, the 3-zone mask does not change the range of linearity. But it has the advantage of reducing non-common path aberrations, because the LOWFS measurement is also sensitive to the position of the opaque disk on the star, while we observed non-common path aberrations with the fully reflective mask.

%%-----------------------------------------------------------

\begin{figure}[b] \centering
\FIG{0.5}{false}{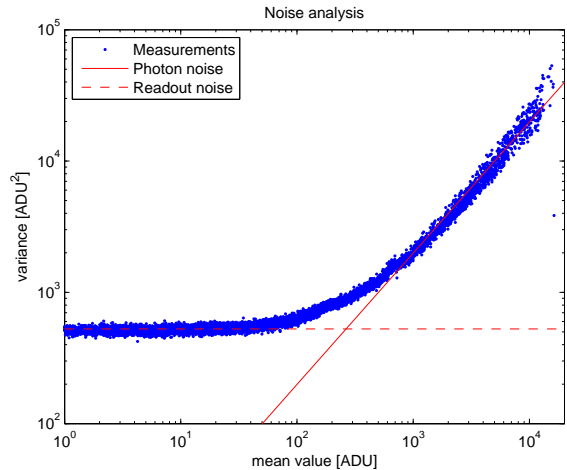}
\caption{Noise analysis of the camera.}
\label{fig:noisevsflux}
\end{figure}

\subsection{Photometry}
\label{sec:Photometry}

To scale our results to future missions like EXCEDE or other \coros, it is important to know the actual flux coming on the camera. So we measured the variance of the noise for different level of flux on each pixels. Figure~\ref{fig:noisevsflux} presents the results: we can clearly see two regimes, the constant readout noise at low flux, and the photon noise at higher flux, where the variance is proportional to the flux. With the best fit of the photon noise, we determined that the gain of the camera is 0.5~photon per ADU.

%%%%%%%%%%%%%%%%%%%%%%%%%%%%%%%%%%%%%%%%%%%%%%%%%%%%%%%%%%%%%

\section{Experimental results}
\label{sec:ExperimentalResults}

%%-----------------------------------------------------------

\subsection{Performances of the LOWFS}
\label{sec:PerformancesOfTheLOWFS}

As explained in Sec.~\ref{sec:HardwareSetup}, the LOWFS has a sampling frequency of $f_\text{samp} = 1.1$~kHz, and we correct only two modes, corresponding to the tip and tilt before the PIAA.

For the first version, we are using a simple integrator law for the control algorithm. The command is delayed by two frames compared to the acquisition, so the rejection function has an overshoot around $f_\text{samp}/10$, with an amplitude dependent of the gain. The problem is that we observe mechanical vibrations around that frequency, so in order to keep those vibrations relatively low, we have to use a small gain of 0.1.

As explained in detail in\cite{Belikov13,Thomas13}, the bench is covered with an enclosure, and the temperature is stabilized at a millikelvin level. Nevertheless, a part of the disturbance we observe is due to residual turbulence inside the enclosure. The rest is due to mechanical vibrations, coming from the bench and the different mounts. Then in open-loop, the tip/tilt disturbance is already low: $5.8\E{-3}$~\loD on the $x$-axis and $9.4\E{-3}$~\loD on the $y$-axis.

With a low gain of 0.1, the cutoff frequency of the integrator law is around $f_\text{samp}/100$, i.e. around 10~Hz. So we are not correcting most of the vibrations, but mostly the turbulence, which is below 1~Hz.

\begin{figure}[b] \centering
  \subfloat[PSD of the LOWFS residue in closed-loop.]
	{\label{fig:vibpsd}\FIG{0.5}{false}{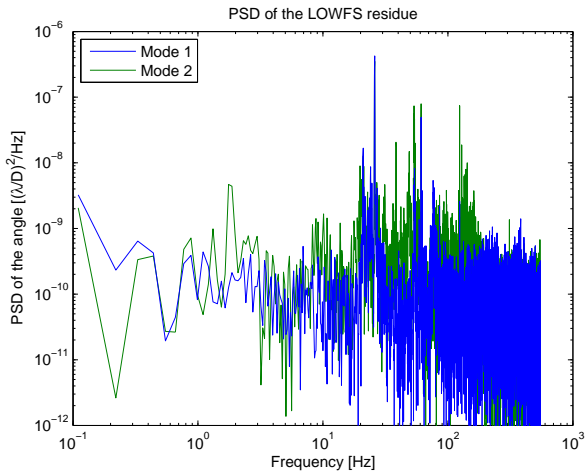}}
  \subfloat[Cumulative integral of the PSD of the residue.]
	{\label{fig:vibint}\FIG{0.5}{false}{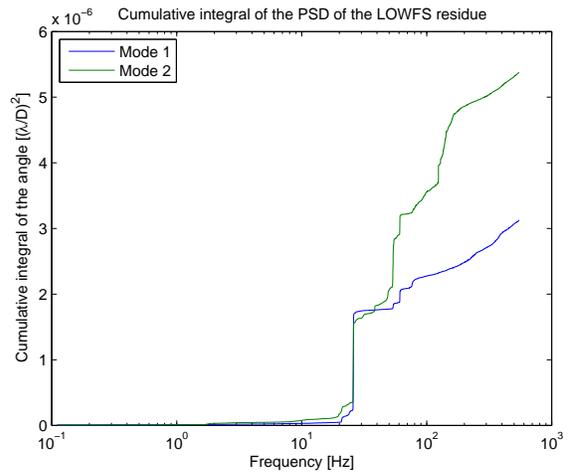}}
  \caption{Vibration analysis with the LOWFS.}
  \label{fig:vibration}
\end{figure}

The results in closed-loop are presented in Fig.~\ref{fig:vibration}: the Power Spectrum Density (PSD) of a 10~s measurement of the residue in Fig.~\ref{fig:vibpsd}, and the cumulative integral of that same PSD in FIg.~\ref{fig:vibint}. In closed-loop, the residue drops to $1.5\E{-3}$~\loD on the $x$-axis, and $2.0\E{-3}$~\loD on the $y$-axis.

On Fig.~\ref{fig:vibpsd}, we can see a lot of vibrations present in the signal:
\begin{itemize}
	\item the electronic frequency at 60~Hz on both axis,
	\item a strong mechanical frequency at 25~Hz on both axis, due to the optical bench,
	\item smaller vibrations at 21, 53 and 75~Hz on the $x$-axis,
	\item smaller vibrations at 2~Hz (probably due to the massive main bench), and also 21, 38, 48, 53 and 122~Hz on the $y$-axis, along with wide contribution between 100 and 200~Hz.
\end{itemize}

The cumulative integral presented in Fig.~\ref{fig:vibint} is very useful to identify the strongest vibrations. Indeed, on this plot, vibrations appear as a sharp rise in the signal. The maximum value in the plot corresponds to the variance of the residue. On this figure, we can see that the frequency at 25~Hz is the strongest, and corresponds to half of the residue in $x$, and a third in $y$. On the $y$-axis, the rest comes mostly from the vibration at 53~Hz, and the wide contribution between 120 and 150~Hz.

To improve the correction, we planned to implement a Linear Quadratic Gaussian controller (LQG) with an automatic identification of the vibrations\cite{Meimon10}. It has already been successfully implemented on the \coro SPHERE\cite{Petit11} and on other experiments\cite{Petit08,Costille11,Lozi11}. With such a controller, we should be able to remove most of the vibrations. Simulations shows an improvement of about $5\E{-4}$~\loD on both axis.

%%-----------------------------------------------------------

\subsection{Characterization of the actuators}
\label{sec:CharacterizationOfTheActuators}

To understand the contribution of the piezo actuators on the tip/tilt measurement, a useful experiment is to measure their step response. Square waves of amplitudes of 0.1 and 0.5~V (0.02~\loD and 0.1~\loD of tip/tilt) are sent to each actuator, with a frequency of 2~Hz, while the LOWFS records measurements with a sample frequency of 1~kHz. By taking an average of the different cycles, we mitigate almost all the disturbance that is not caused by the actuators, and keep only their response.

\begin{figure}[b] \centering
  \subfloat[Response on the $x$-axis.]
	{\label{fig:response1}\FIG{0.5}{false}{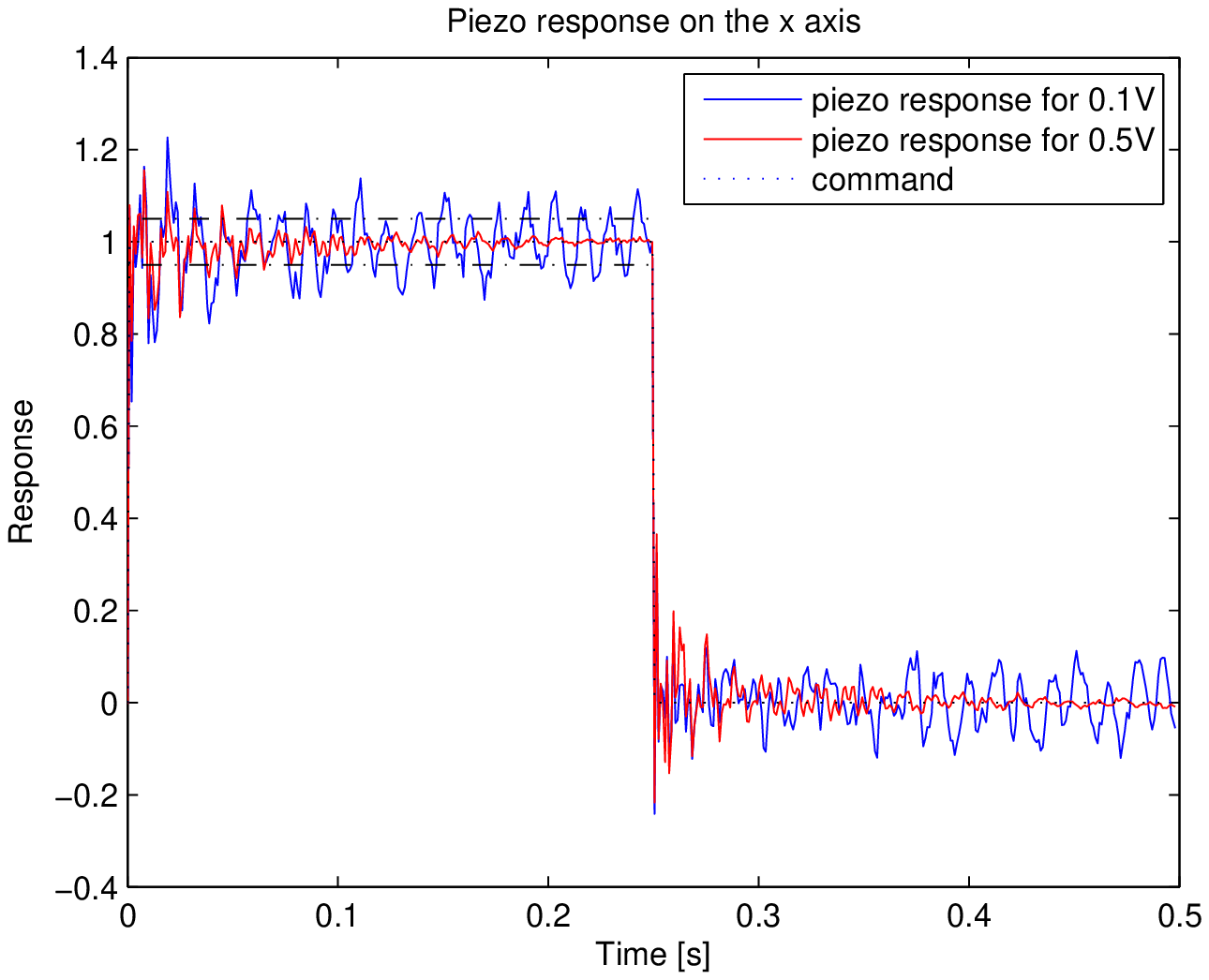}}
  \subfloat[Response on the $y$-axis.]
	{\label{fig:response2}\FIG{0.5}{false}{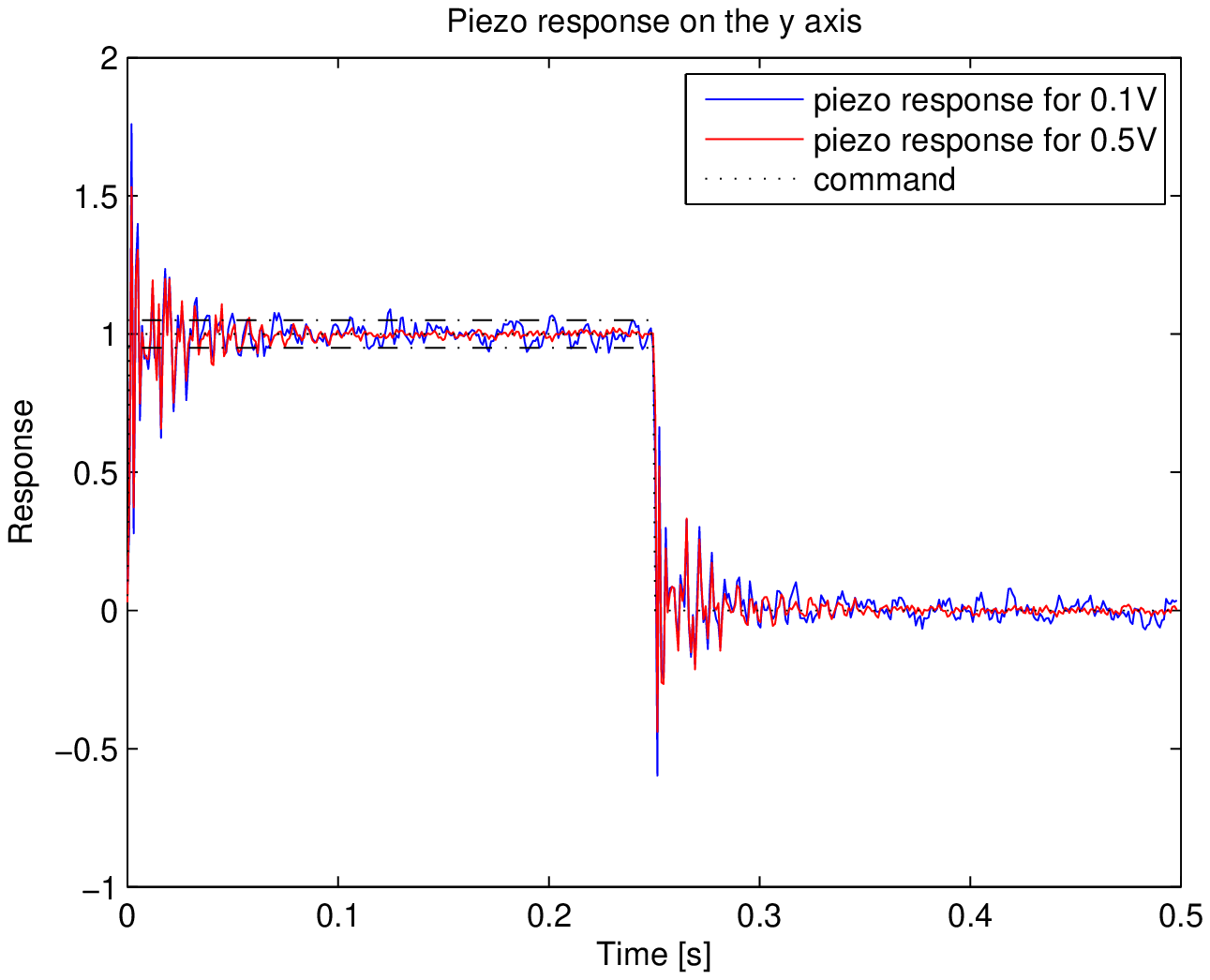}}
  \caption{Normalized response of the piezoelectric actuators.}
  \label{fig:response}
\end{figure}

Figure~\ref{fig:response} presents the results of this experiment. On this figure, we can see that the response time is very fast, around 1~ms, i.e. the frame period of the LOWFS. But the response suffers of a high overshoot that takes time to settle down, especially on the $y$-axis. It has clearly the same relative amplitude for 0.1~V and 0.5~V: around 25\% on the $x$-axis, and 75\% on the $y$-axis. But measurements at 0.1~V are more noisy, especially on the $y$-axis; this corresponds to a residual vibration that is not due to the actuators because it is dependent of the voltage. Then the real response of the piezo actuators is closer to the red plots than the blue plots. Also, on both axes, the settling time at $\pm5\%$ is close to 50~ms. The design is then compatible with a control at 1~kHz, but the piezo actuators are introducing some vibrations.

\begin{figure} \centering
\FIG{0.5}{false}{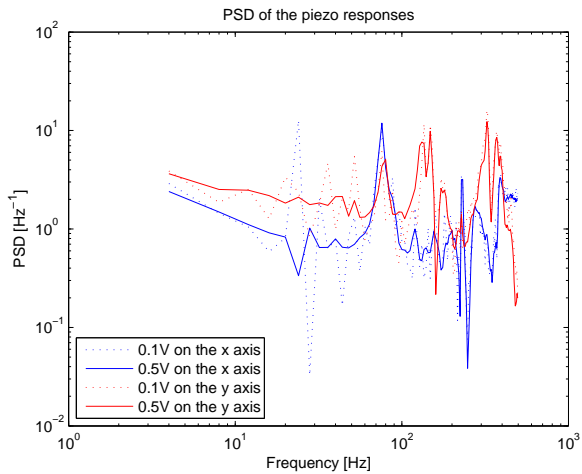}
\caption{PSD of the normalized responses.}
\label{fig:psdresponse}
\end{figure}

The figure~\ref{fig:psdresponse} presents the PSD of the responses. We can see that the $x$-axis is dominated by a vibration at 80~Hz, and smaller contributions at 220~Hz and 400~Hz. The $y$-axis is clearly more noisy, with contributions at 80, 120, 150, 350 and 400~Hz. The common contribution at 80~Hz seems to have the same source, it might corresponds to the frequency of the whole mount on which the fiber head is attached. This frequency is also visible in the LOWFS measurements presented in Sec.~\ref{sec:PerformancesOfTheLOWFS}, but it is not the predominant component. On the $y$-axis, we can also see the contributions at 120 and 150~Hz in the LOWFS measurements.

%%-----------------------------------------------------------

\subsection{Impact of the flux}
\label{sec:ImpactOfTheFlux}

The precision of the LOWFS depends on different factors:
\begin{itemize}
	\item the intensity of the source,
	\item the sampling of the camera,
	\item the bit depth of the pixels,
	\item the size of the opaque disk on the focal plane mask,
	\item the defocus applied to the camera,
	\item other factors (readout noise, exposure, alignment, etc.)
\end{itemize}

Here, we are analyzing the impact of the flux of the source on the residue. Actually, we modified the number of photons not by changing the intensity of the source, but by varying the exposure time of the camera.

For each exposure time, we analyzed the LOWFS measurement and the centroid, and calculated two quantities:
\begin{itemize}
	\item the residue, which is simply the standard deviation of the measurement,
	\item the measurement noise, calculated by extrapolating the noise between 400~Hz and $f_\text{samp}/2=550$~Hz in the PSD.
\end{itemize}

We also compared two thresholds applied to the images: no threshold and a threshold at 200~ADU, corresponding to 1.2\% of the saturation level.

\begin{figure} \centering
  \subfloat[Measurements where no threshold is applied to the images.]
	{\label{fig:noisevsexp1}\FIG{0.485}{false}{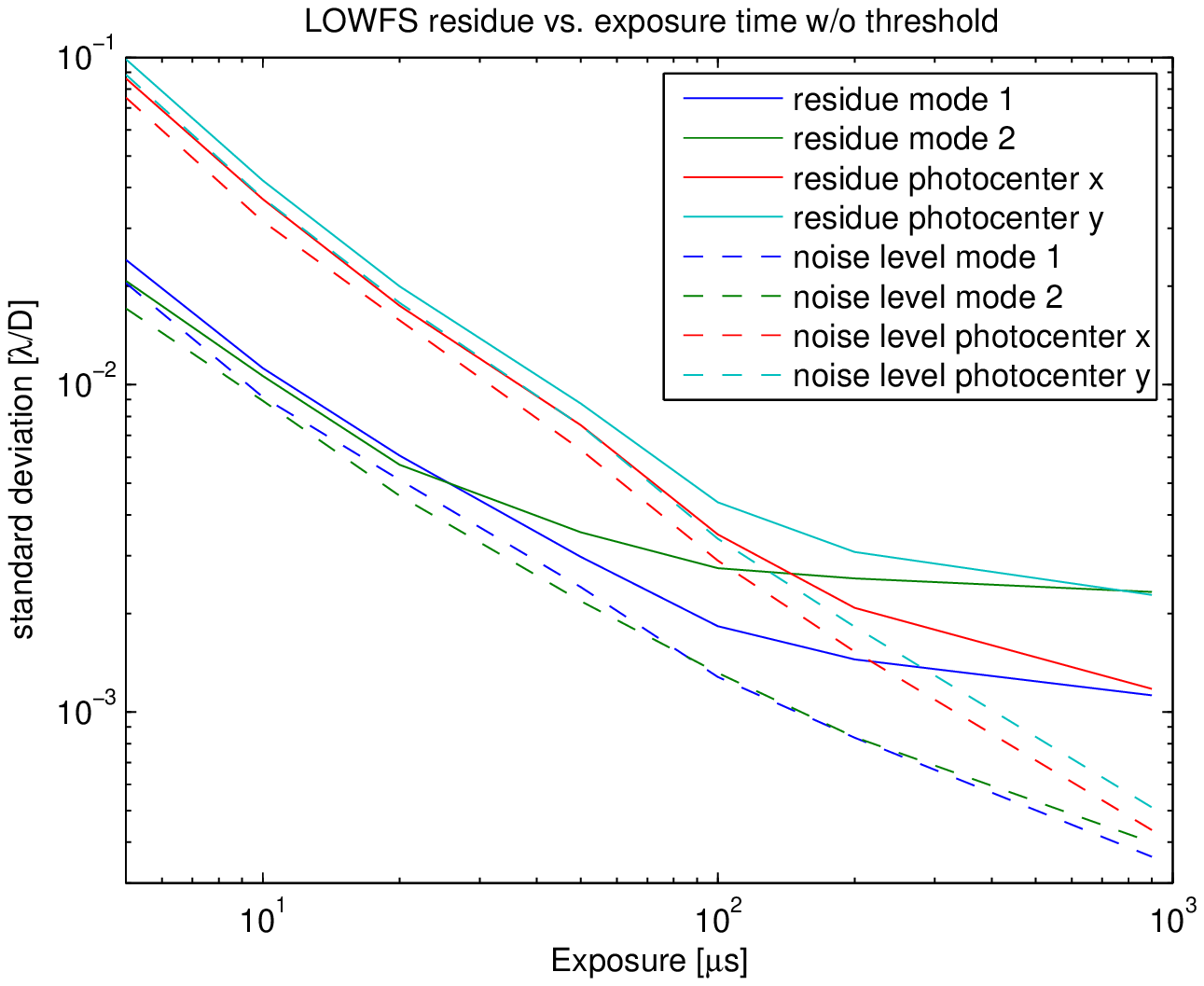}}\hspace{10pt}
  \subfloat[Measurements where a fixed threshold is applied at 200~ADU.]
	{\label{fig:noisevsexp2}\FIG{0.485}{false}{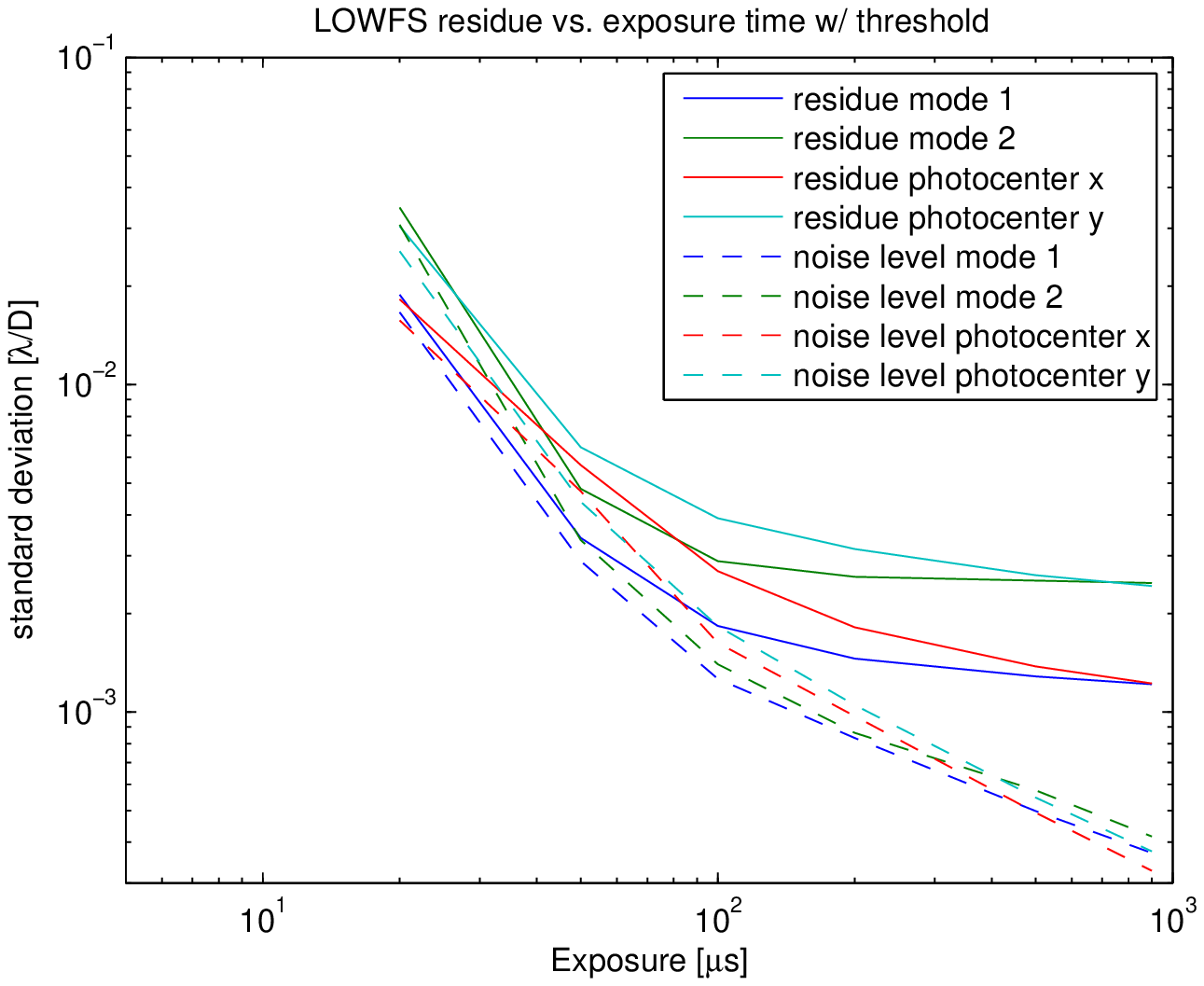}}
  \caption{LOWFS residue and measurement noise for different exposure times.}
  \label{fig:noisevsexp}
\end{figure}

Results are presented in Fig.~\ref{fig:noisevsexp}: Fig.~\ref{fig:noisevsexp1} is without a threshold, and Fig.~\ref{fig:noisevsexp2} is with the threshold at 200~ADU.

In both figures, the residue seems to be limited by the noise level only for exposures below 100~\mus. For higher exposures, the residue is almost stable, limited only by the vibrations. With or without a threshold, we can see that the method used by the LOWFS is less noisy than a simple centroid.

Without a threshold, the noise level of the centroid measurement is proportional to the inverse of the number of photons: it is limited by the readout noise. As for the LOWFS, it is limited by photon noise over 100~\mus, and readout noise below.

The fixed threshold does not help the LOWFS: it is still limited by photon noise over 100~\mus, but the residue is proportional to the inverse square of the number of photons. On the contrary, it helps the centroid measurement, because the noise level is almost the same as for the LOWFS.

This study showed that the LOWFS we implemented, even if we correct only two modes, is more efficient than a simple centroid, especially when the number of photons is low. Of course this method has other advantages, because we can correct more than just tip and tilt.

By scaling those results to the EXCEDE mission or other future \coros, we can deduce the expected performances. A few other studies has to be performed, for example on the influence of the sampling of the camera.

%%-----------------------------------------------------------

\subsection{Impact of the wavefront control}
\label{sec:ImpactOfTheWavefrontControl}

With a wavefront control ---~either Electric Field Conjugation (EFC) or speckle nulling\cite{Thomas13}~---, the shape of the PSF changes a lot, because the light inside the dark zone is shifted to the other side of the PSF. This change of shape affects the measurement of the LOWFS, because it can be considered as a tip/tilt error by the controller.

\begin{figure} \centering
  \subfloat[Profile of the PSF during the first seven iterations of EFC.]
	{\label{fig:psfwwfc}\FIG{0.485}{false}{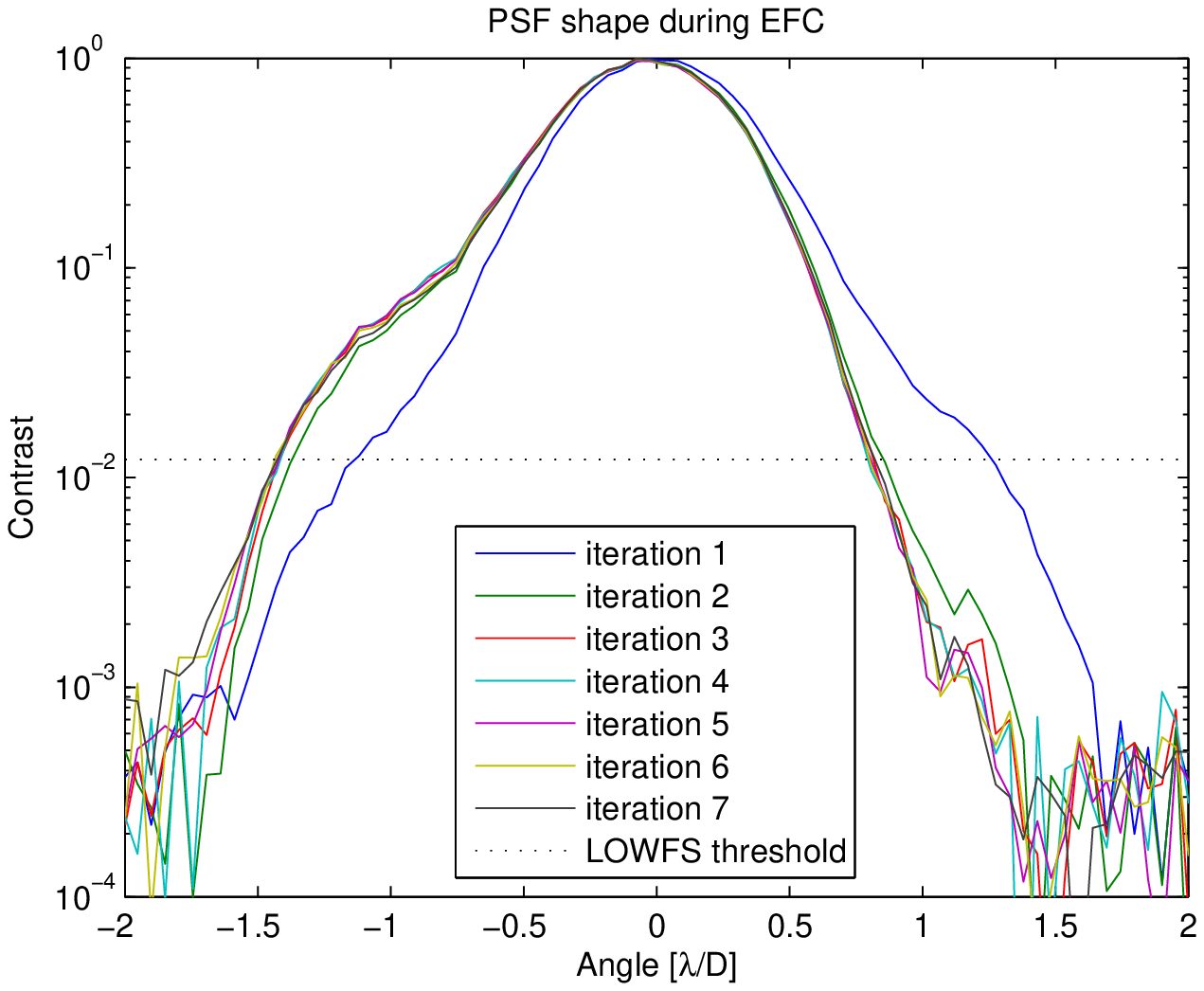}}\hspace{10pt}
  \subfloat[Measurement of the LOWFS in open loop during EFC.]
	{\label{fig:lowfswwfc}\FIG{0.485}{false}{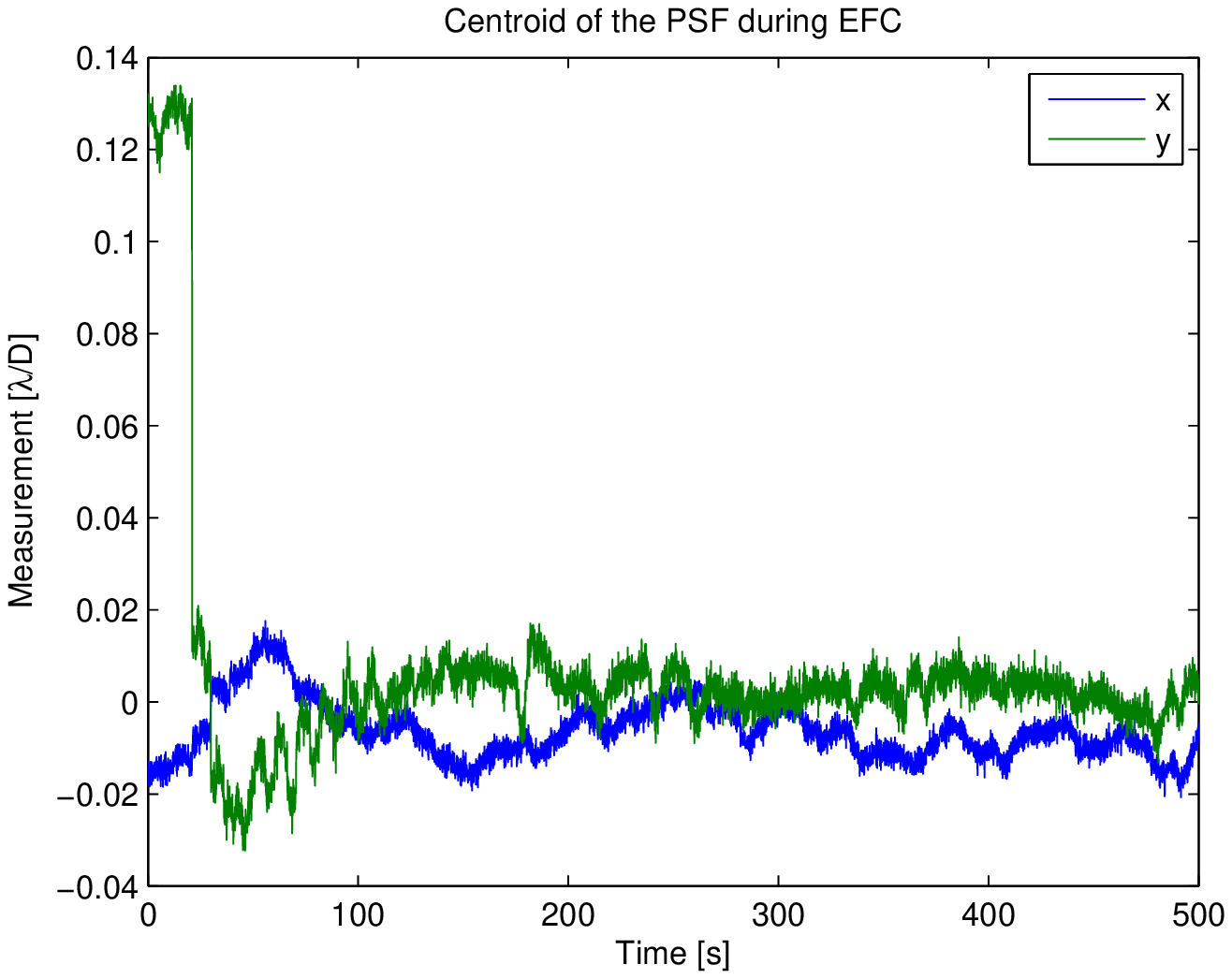}}
  \caption{Impact of the wavefront control on the PSF and the LOWFS.}
  \label{fig:wfc}
\end{figure}

Figure~\ref{fig:psfwwfc} presents the PSF profiles during the first seven iterations of EFC, taken with the science camera. Those profiles are along the $y$-axis of the bench, in the middle of the dark zone. On the first iteration, the DM is flat, and no dark zone is created. The PSF is then almost symmetric, with a contrast higher than $\e{-2}$ at 1.2~\loD.

After the first shape is sent to the DM, the contrast in the dark zone drops lower than $\e{-3}$, and after the third iteration, the modifications of shapes are below the readout noise of the science camera.

In Fig.~\ref{fig:psfwwfc}, the dotted line corresponds to a threshold applied to the LOWFS images, to remove the noise in dark areas. We can see that after iteration~3, the shape of the PSF over the threshold does not change anymore.

This is confirmed by Fig.~\ref{fig:lowfswwfc}, corresponding to the LOWFS measurement during EFC. on the $y$-axis, we can see a first step at around 20~s, corresponding to the difference between iterations~1 and~2, and a second smaller step between iterations~2 and~3 at around 30~s. After those steps, the variations we observe correspond to other contributions, i.e. turbulence and vibrations. So the LOWFS is not affected by PSF changes after iteration~3.

The difference between a flat DM and a dark zone shifts the measurement of about 0.15~\loD, which corresponds to the correction we have to apply on the focal plane mask to keep the IWA at 1.2~\loD.

So the LOWFS should be used only after a few iterations of the WFC, to avoid any residual tip/tilt due to the correction. In any case, in the first iterations, the contrast is high enough that the system is not sensitive to tip/tilt errors, it is only useful when the contrast reaches at least $\e{-6}$ at 1.2~\loD.

%%-----------------------------------------------------------

\subsection{Contrast sensitivity to tip/tilt errors}
\label{sec:SensitivityOfTheContrast}

To understand if tip/tilt errors are limiting us on the bench, we injected different levels of tip/tilt errors, and we measured the contrast in two regions of interest: an inner zone between 1.2 and 2~\loD, and an outer zone between 2 and 4~\loD.

The injected noise is a random white noise with an adjustable amplitude, sent on the two piezo actuators controlling the position of the fiber. We measured each time the real amplitude of the noise, because it depends also on the disturbances already present on the bench, the response of the actuators, which modifies the amplitude of the noise at high frequency, and the response of the controller, which reduces the injected noise at low frequencies.

\begin{figure}[b] \centering
\FIG{0.5}{false}{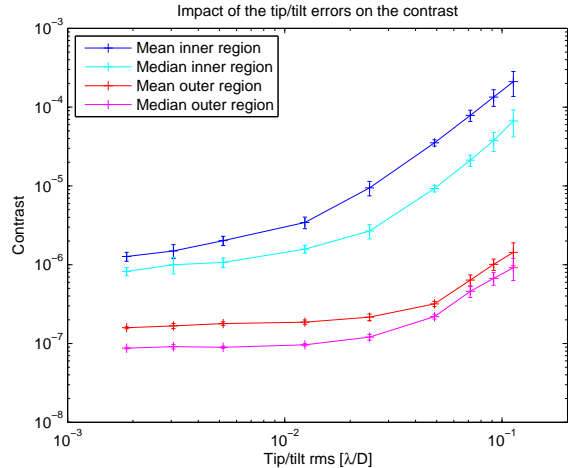}
\caption{Contrast at different levels of tip/tilt errors.}
\label{fig:contvstt}
\end{figure}

Figure~\ref{fig:contvstt} presents the results of that experiment. The initial contrast without injected noise was better than our goal: $8\E{-7}$ in the inner region, and $9\E{-8}$ in the outer region.

In this figure we can see two regimes:
\begin{itemize}
	\item The contrast is almost not affected by tip/tilt errors below $\e{-2}$~\loD rms in the inner region, and $3\E{-2}$~\loD rms in the outer region.
	\item The contrast is proportional to the square of the tip/tilt after $2\E{-2}$~\loD rms in the inner region, and around $5\E{-2}$~\loD rms in the outer region.
\end{itemize}

Thanks to this experiment, we deduced that we can accept tip/tilt errors up to almost $\e{-2}$~\loD rms and still meet our goal of a $\e{-6}$ contrast in the inner region. Also, it is clear that the tip/tilt errors are no more limiting our result. Since that experiment, we obtained a better contrast at $4\E{-7}$\cite{Thomas13}. At this level of contrast, it is more sensitive to tip/tilt errors.

%%%%%%%%%%%%%%%%%%%%%%%%%%%%%%%%%%%%%%%%%%%%%%%%%%%%%%%%%%%%%

\section{Conclusion}
\label{sec:Conclusion}

This paper presented a successful implementation of a coronagraphic low-order wavefront sensor. With a simple design, we were able to reduce pre-PIAA tip/tilt motions from between 5 and $9\E{-3}$~\loD rms in open-loop to $\approx2\E{-3}$~\loD rms in closed-loop. With a frame rate of 1.1~kHz, it was also very useful to perform a vibration analysis, leading to a modification of the most noisy elements. We also demonstrated that for the level of contrast and inner working angle we want for the EXCEDE mission, tip/tilt errors are currently not the limiting factor for our results.

This LOWFS is very interesting for any type of \coro using a focal plane mask, because its design is very simple, and it can be rapidly implemented on any testbench or instrument. The results we presented here can be scaled to any future mission, for example EXCEDE or AFTA, even if they do not use a PIAA to apodize the beam.

In future developments, we will implement a linear quadratic Gaussian controller, based on a Kalman filter, to remove residual vibrations. We also want to inject typical vibrations of a space mission to be as realistic as possible for our tests. The LOWFS will also soon be tested in vacuum on a testbench we are building at Lockheed Martin in Palo Alto.

%%%%%%%%%%%%%%%%%%%%%%%%%%%%%%%%%%%%%%%%%%%%%%%%%%%%%%%%%%%%%
%\acknowledgments     %>>>> equivalent to \section*{ACKNOWLEDGMENTS}       

%%%%%%%%%%%%%%%%%%%%%%%%%%%%%%%%%%%%%%%%%%%%%%%%%%%%%%%%%%%%%
%%%%% References %%%%%

\bibliography{report}   %>>>> bibliography data in report.bib
\bibliographystyle{spiebib}   %>>>> makes bibtex use spiebib.bst

\end{document}